\documentclass[10pt,twocolumn,letterpaper]{article}

\usepackage{cvpr}
\usepackage{times}
\usepackage{epsfig}
\usepackage{graphicx}
\usepackage{amsmath}
\usepackage{amssymb}
\usepackage{mwe}
\usepackage{adjustbox}
\usepackage{subcaption}
\usepackage{booktabs}
\usepackage[dvipsnames,svgnames,x11names]{xcolor}
\usepackage{cuted}
\usepackage{capt-of}
\usepackage{kotex}

\usepackage{multirow}

\newcommand{\myparagraph}[1]{\vspace{2pt}\noindent{\bf #1}}

\usepackage[pagebackref=true,breaklinks=true,letterpaper=true,colorlinks,bookmarks=false]{hyperref}
\cvprfinalcopy

\DeclareMathOperator*{\argmin}{argmin}

\begin{document}
\title{Sound-Guided Semantic Image Manipulation}

 \author{
Seung Hyun Lee$^1$, Wonseok Roh$^1$, 
Wonmin Byeon$^4$, 
Sang Ho Yoon$^{3}$,
Chanyoung Kim$^1$,\\
  Jinkyu Kim$^{2*}$, and Sangpil Kim$^{1}$\thanks{Corresponding authors.}\\
  $^1$Department of Artificial Intelligence, Korea University\\
  $^2$Department of Computer Science and Engineering, Korea University\\
  $^3$Graduate School of Culture Technology, KAIST\\
  $^4$NVIDIA Research, NVIDIA Corporation\\
}

\maketitle

\begin{strip}\centering
\vspace{-1.59cm}
\includegraphics[width=\textwidth]{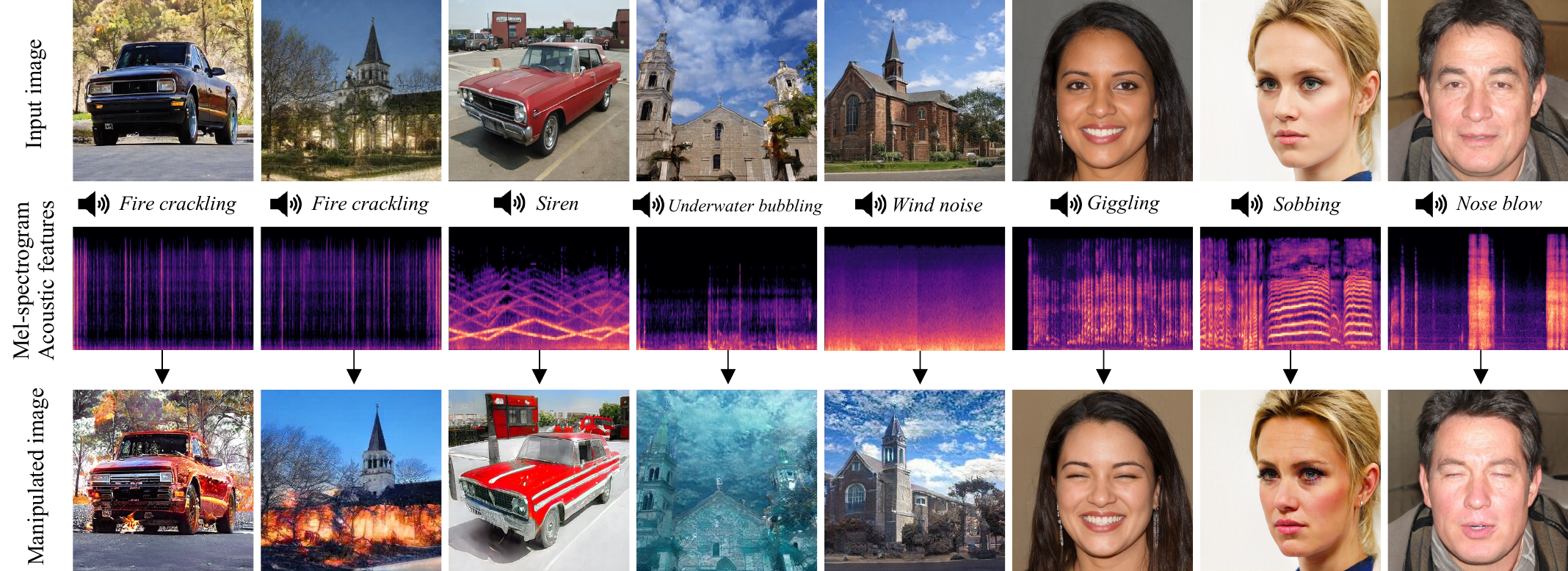}
\captionof{figure}
{
Modified images with sound-guided semantic image manipulation.
Our method manipulates source images~(top row) given user-provided sound~(middle row) into semantic images~(last row).
}\label{fig:fig1}\vspace{-0.5em}
\end{strip}

\begin{abstract}
\vspace{-1em}
The recent success of the generative model shows that leveraging the multi-modal embedding space can manipulate an image using text information. However, manipulating an image with other sources rather than text, such as sound, is not easy due to the dynamic characteristics of the sources. Especially, sound can convey vivid emotions and dynamic expressions of the real world. Here, we propose a framework that directly encodes sound into the multi-modal~(image-text) embedding space and manipulates an image from the space. Our audio encoder is trained to produce a latent representation from an audio input, which is forced to be aligned with image and text representations in the multi-modal embedding space. We use a direct latent optimization method based on aligned embeddings for sound-guided image manipulation.
We also show that our method can mix different modalities, i.e., text and audio, which enrich the variety of the image modification. The experiments on zero-shot audio classification and semantic-level image classification show that our proposed model outperforms other text and sound-guided state-of-the-art methods. 

\end{abstract}


\section{Introduction}
\vspace{-0.5em}


Image editing has been widely studied in the field of computer vision due to 
its usefulness in photo-realistic editing applications, social media image sharing, and image-based advertisement.
An image can be used to transfer its style into the target image~\cite{gatys2016image, gatys2016neural}.
Also, modifying specific parts in the human face image, such as hairstyle or color, is useful in image editing applications~\cite{xia2021tedigan,Patashnik_2021_ICCV}.
The purpose of semantic image manipulation is to generate a novel image that contains both source image identification and semantic information of user intention. 
In this paper, we tackle the semantic image manipulation task, which is the task of modifying an image with user-provided semantic cues.
To apply the user intention into the image, a mixture of sketches and text is used to perform image manipulation and synthesis~\cite{park2019semantic,xia2021tedigan}.
User intention can be applied by drawing a paint~\cite{park2019semantic} or writing text with semantic meanings~\cite{xia2021tedigan,gatys2016neural}.

Text-based image manipulation methods are proposed to edit the image conditionally~\cite{el2019tell, jiang2021language, li2020manigan, nam2018tagan,xia2021tedigan}.
These works modify target contents in the image based on the text information.
Among the text-based image manipulation methods, StyleCLIP~\cite{Patashnik_2021_ICCV} considered leveraging the representational power of Contrastive Language-Image Pre-training (CLIP)~\cite{radford2learning} models to produce text-relevant manipulations with given text input. StyleCLIP maintains high quality image generation ability using StyleGAN~\cite{jeong2021tr} while allowing insertion of semantic text into the image. 
However, text-based image manipulation has an inherent limitation when applying sound semantics into the image, 
due to the lack of handling vivid sound, which has infinite variation.
Since the text is the form of discrete character, expressing the spectrum that has a continuous and dynamic context of sound is extremely difficult in our world. 
For example, every ``thunder'' generates different loudness and characteristic of ``\textit{sound of thunder}''.
The discreetness of a text message prevent expressing the detailed difference of the sound around us.
Therefore, the text-based image manipulation model has limitations in transferring specific, vivid sound semantics into the source image for the image modification. 

Sound provides polyphonic information of the scene and contains multiple sound events~\cite{9524590}.
That is why watching a movie with sound is more realistic than reading a book.
Our daily environment is filled with diverse sound sources and a complex blend of audio signals~\cite{9524590}.
Therefore, sound, which we focus on, is a necessary modality for image manipulation.
 
 Several studies~\cite{chen2017deep, hao2018cmcgan, oh2019speech2face, qiu2018image, wan2019towards, zhu2021deep} have attempted to visualize the meaning of sound, but it is still challenging to reflect sound events in high-resolution images due to two reasons. 
 The first reason is the lack of a suitable high-resolution audio-visual dataset. Audio-visual benchmark video datasets~\cite{caba2015activitynet, kay2017kinetics, soomro2012ucf101} for GAN training has generally lower resolution than high-resolution image datasets including Flickr-Faces-HQ (FFHQ)~\cite{karras2019style} and The Large-scale Scene Understanding Challenge (LSUN)~\cite{yu2015lsun}. There is no dataset with as many audio-visual pairs as the number of image-text pairs used for CLIP training. CLIP uses 400 million image-text pair data to learn the relationship between very large and diverse image and text modalities, whereas audio-visual pair data is still insufficient. Secondly, it is difficult to discover potential correlations between auditory and visual modalities~\cite{zhu2021deep}. 
Extracting appropriate temporal context, tone, and theme from the sound is difficult.

\begin{figure*}
    \begin{center}
        \includegraphics[width=\textwidth]{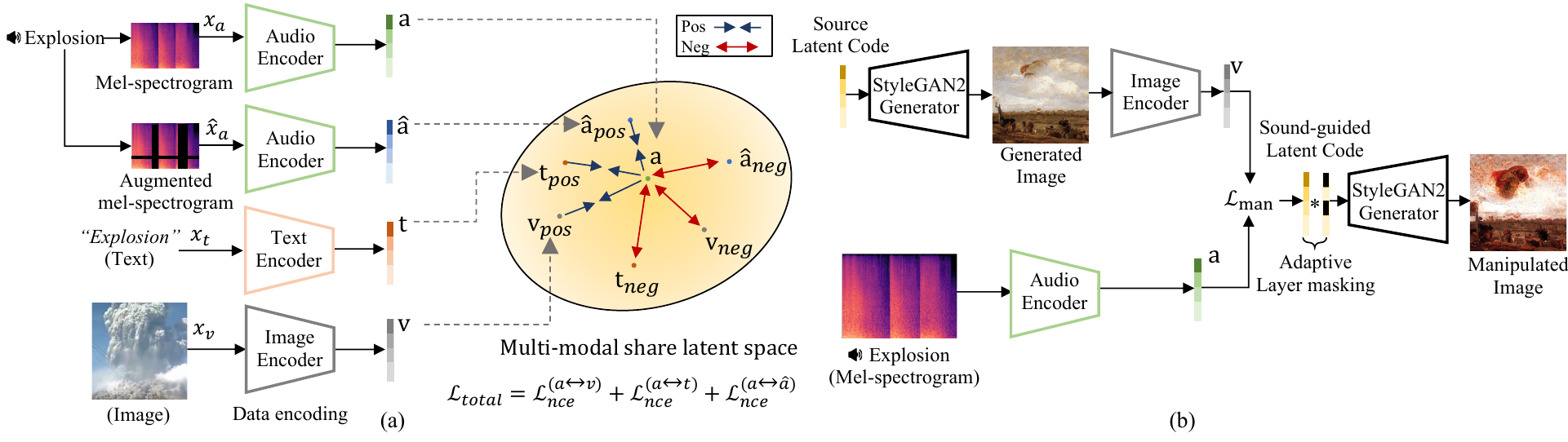}
    \end{center}
    \vspace{-1.3em}
    \caption{Our model consists of two main steps: (a) the {\em CLIP-based Contrastive Latent Representation Learning} step and (b) the {\em Sound-Guided Image Manipulation} step. In (a), we train a set of encoders with three different modalities~(audio, text, and image) to produce the matched latent representations. The latent representations for a positive triplet pair~(e.g., audio input: ``Explosion'', text: ``explosion'', and corresponding image) are mapped close together, while that of negative pair samples further away in the~(CLIP-based) embedding space~(left).  In (b), we use a direct code optimization approach where a source latent code is modified in response to user-provided audio, producing a sound-guided image manipulation result~(right).}
    \label{fig:contrastivelearning}
    \vspace{-1.3em}
\end{figure*}

To overcome these challenges of manipulating images with sound semantics, we introduce a novel image manipulation method driven by sound semantics~(see Fig.~\ref{fig:contrastivelearning}). As shown in Fig.~\ref{fig:fig1}, an image of an old car is manipulated into an old car with a fire truck-like exterior appearance when adding a siren sound. 
Our model consists of two main stages: (i) the CLIP-based Multi-modal Representation Learning, where an audio encoder is trained to produce a latent representation aligned with textual and visual semantics by leveraging the representation power of pre-trained CLIP models.
(ii) the Sound-Guided Image Manipulation, where we use the direct latent code optimization to produce a semantically meaningful image in response to a user-provided sound.


Our experimental results show that the proposed method supports a variety of sound sources with a better reflection of given audio information when transferring image styles. The sound-based approach supports more diverse and detailed information related to scenes compared to text-based image manipulation methods. 





Our main contributions are listed as follows: 
\begin{itemize}
    \vspace{-0.7em}
    \item We propose multi-modal contrastive losses to expand the CLIP-based embedding space. Moreover, we introduce contrastive learning on augmented audio data, which helps to learn a more robust representation. Here, we achieve state-of-the-art performance for a zero-shot audio classification task.\vspace{-0.7em}
    \item We propose semantic-level image manipulation solely based on the given audio features, including temporal context, tone, and volume.\vspace{-0.7em}
    \item We propose the sound-guided code optimization steps with adaptive layer masking for putting sound meaning into images, enhancing the realism of the output.\vspace{-0.7em}
\end{itemize}

\section{Related Work}

\myparagraph{Text-guided Image Manipulation.} Text-guided image manipulation is the most widely studied among guidance based tasks.
Several studies~\cite{dong2017semantic, li2020manigan, nam2018tagan} employed the GAN-based encoder-decoder structure to preserve the features of the image while presenting image manipulations corresponding to the text description. 
StyleCLIP~\cite{Patashnik_2021_ICCV} and TediGAN~\cite{xia2021tedigan} utilize the latent space of the pre-trained StyleGAN and the prior knowledge from CLIP~\cite{radford2learning}. StyleCLIP performed image manipulation using a user-provided text prompt. TediGAN enabled image generation and manipulation using GAN inversion technique using multi-modal mapping. Beyond text and images, a sound can express a complex context appearing in a scene, and there is correspondence between a sound and an event occurring in the scene.

\myparagraph{Sound-guided Image Manipulation.} Sound contains temporal dynamic information of a scene, which can be used as an imagery source for image manipulation. 
Some approaches have been introduced for the sound-guided image manipulation task.
However, the previous works mainly focus on music (instead of using sound semantics), which includes music-to-visual style transfer with cross-modal learning strategy~\cite{lee2020crossing} and a neural music visualizer by mapping music embeddings to visual embeddings from StyleGAN~\cite{jeong2021tr}. To manipulate the image according to the sound, 
\textit{Tr$\ddot{a}$umerAI}~\cite{jeong2021tr} visually expresses music by latent transfer mapping of music to StyleGAN's style embedding. 

However, the above studies have limitations in focusing only on the reaction, not the semantic of the sound, in the direction of navigation in the latent space of StyleGAN. \textit{Crossing you in style}~\cite{lee2020crossing} uses the period to define the semantic relationship between sound and visual domain, but there is still a limitation which only can transfer the image style.
Our proposed method can isolate the modification area in the source image, such as modifying the emotion of the face while preserving the color of the hair.


\myparagraph{Interpreting Latent Space in StyleGAN.} The intermediate latent space in pre-trained StyleGAN~\cite{karras2019style} solves the disentanglement issue and allows the generated images to be manipulated meaningfully according to changes in the latent space. 
Extended latent space $\mathcal{W}+$ allows image manipulation with interpretable controls from a pre-trained GAN generator~\cite{abdal2019image2stylegan, karras2019style, karras2020analyzing}. 
For latent space analysis in audio sequences, \textit{Audio-reactive StyleGAN}~\cite{brouwer2020audio} generates an image every time step by calculating the magnitude of the audio signal and moving it in the latent space of StyleGAN. However, the method cannot control the meaning of sound in the latent space. StyleGAN's motion in the latent space is only mapped to the magnitude of the sound. There is a novelty in that we manipulate images with the properties of sound. 

\myparagraph{Audio-visual Representation Learning.} Cross-modal representation learning obtains relationships between different modalities in audio-visual tasks such as video retrieval and text-image cross-modal tasks such as image captioning and visual question answering. Audio-visual representation learning studies~\cite{DBLP:journals/corr/AytarVT17,nagrani2018learnable, suris2018cross} aim to map both modalities to the same embedding space. The correlation between modalities is learned by contrastive learning between composite audio-visual pairs~\cite{chen2021distilling, mazumder2021avgzslnet, sun2020learning}. 

However, audio-visual representation learning is still challenging because there is no adequate data as much as CLIP~\cite{radford2learning} for learning the correlation between different modalities.
CLIP learned the relationship between image and text embedding by multi-modal self-supervised learning of 400 million image-text pairs and showed zero-shot inference performance comparable to supervised learning in most image-text benchmark datasets.  

In this paper, the audio encoder not only exploits the representation ability of CLIP but also learns supervisory signals from the audio data itself through self-supervised manners. As a result, our method obtains an audio-specific representation for sound-guided image manipulation.


\section{Method}
We follow the existing text-guided image manipulation model, StyleCLIP~\cite{Patashnik_2021_ICCV}. Our model and StyleCLIP manipulate the latent code of StyleGAN using joint embedding space between modalities. However, our model extends the CLIP~\cite{radford2learning} embedding space to the audio embedding space, which was not embedded before. We also introduce novel contrastive losses and adaptive masking for sound-guided image manipulation.
Our model consists of two main steps: (i)~the CLIP-based Multi-modal Latent Representation Learning and (ii)~the Sound-guided Image Manipulation. First, we train audio, text, and image encoders to generate new latent representations. To do so, we train the audio encoder using the InfoNCE loss~\cite{oord2018representation, alayrac2020self, zhang2020contrastive} to produce a latent representation that is aligned with the representations from the pre-trained CLIP's text and image encoders. Such aligned representations can be used for image manipulation with the provided audio input. After the pre-training step, we use encoders to manipulate images according to a target sound input~(e.g., images with different facial expressions can be manipulated with different sound inputs).

\subsection{Multi-modal Latent Representation Learning}
As shown in Fig.~\ref{fig:contrastivelearning}~(a), we train a set of encoders with three different modalities \{audio, text, and image\} to produce the matched representations in the embedding space. Specifically, given audio, text, and image inputs, i.e. $x_a$, $x_t$, and $x_v$, we use three different encoders to obtain a set of $d$-dimensional latent representations, i.e. $\bf{a}$, $\bf{t}$, and ${\bf{v}}\in\mathcal{R}^{d}$, respectively. These latent representations are learned via a typical contrastive learning approach following the work by Radford~\etal~\cite{radford2learning} -- the latent representations for a positive triplet pair are mapped close together in the embedding space, while that of negative pair samples further away. Learning such a joint representation from scratch is, however, generally challenging due to the lack of multi-modal datasets, which can provide positive and negative pairs. Thus, we instead leverage the pre-trained CLIP model, which optimized a visual-textual joint representation by contrastive learning. Then, we train an audio encoder to produce an aligned representation by using contrastive learning. Details are explained in the next section. Note that we obtain a latent representation $\hat{{\bf{a}}}\in\mathcal{R}^{d}$ from an augmented audio input $\hat{x}_a$, which is shown useful to improve the quality of the latent representation as this is a common practice in the self-supervised representation learning. 




\myparagraph{Matching Multi-modal Representations via Contrastive Loss.}
We use the InfoNCE loss~\cite{alayrac2020self} to map positive audio-text pairs close together in the CLIP-based joint embedding space, while negative pairs further away. Formally, given a minibatch of $N$ audio-image representation pairs $\{{\bf{a}}_i, {\bf{t}}_j\}$ for $i\in\{1, 2, \dots, N\}$, we first compute the following audio-to-text loss function for the $i$-th pair:
\begin{equation}
    l_{i}^{(a\rightarrow t)}=-\text{log}\cfrac{\exp(\langle{\bf{a}}_i, {\bf{t}}_j\rangle/\tau) }{\sum_{\textnormal{j=1}}^N\exp(\langle{\bf{a}}_i, {\bf{t}}_j\rangle/\tau)}
    \label{loss:mini_con}
\end{equation}
where $\langle{\bf{a}}_i, {\bf{t}}_j\rangle$ represents the cosine similarity, i.e. $\langle{\bf{a}}_i, {\bf{t}}_j\rangle = {\bf{a}}_i^\intercal{\bf{t}}_j/\|{\bf{a}}_i\|\|{\bf{t}}_j\|$ and $\tau$ is a temperature parameter. This loss function is the log loss of an $N$-way classifier that wants to predict $\{{\bf{a}}_i, {\bf{t}}_j\}$ as the true representation pair. As the loss function is asymmetric, we define the following similar text-to-audio contrastive loss:
\begin{equation}
    l_{i}^{(t\rightarrow a)}=-\text{log}\cfrac{\exp(\langle{\bf{t}}_i, {\bf{a}}_j\rangle/\tau) }{\sum_{\textnormal{j=1}}^N\exp(\langle{\bf{t}}_i, {\bf{a}}_j\rangle/\tau)}
    \label{loss:mini_con2}
\end{equation}
Concretely, we minimize the following loss function $\mathcal{L}_{\textnormal{nce}}$ as a sum of the two losses $l_{i}^{(a\rightarrow t)}$ and $l_{i}^{(t\rightarrow a)}$ for all positive audio-text representation pairs in each minibatch of size $N$:
\begin{equation}
    \begin{aligned}
    \mathcal{L}_{\textnormal{nce}}^{(a \leftrightarrow t)}=\cfrac{1}{N}\sum_{i=1}^N (l_{i}^{(a\rightarrow t)} + l_{i}^{(t\rightarrow a)})
    \end{aligned}
    \label{loss:nce}
\end{equation}

\begin{figure}[t]
  \centering
  \includegraphics[width=\linewidth]{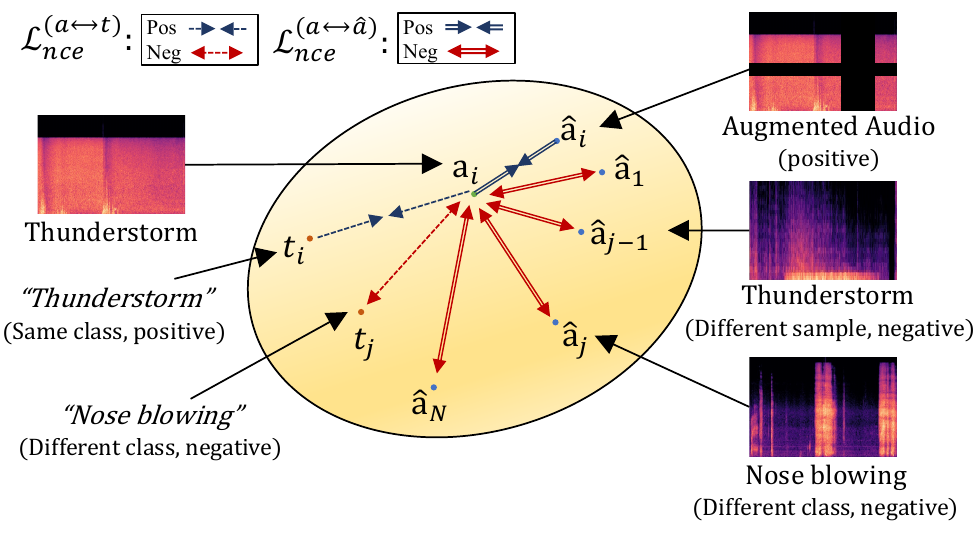}
  \vspace{-1.3em}
  \caption{Multi-modal contrastive learning with audio self-supervised loss. 
  }
  \vspace{-1.3em}
  \label{fig:new_figure}
\end{figure}

\myparagraph{Applying Self-supervised Representation Learning for Audio Inputs.}
Self-supervised learning approaches rely on a contrastive loss that encourages representations of the same-class different views to be close in the embedding space, while that of different-class views to be pushed away from each other. We apply this technique to improve the quality of audio representations by minimizing the following $\mathcal{L}_{\textnormal{self}}^{(a \leftrightarrow \hat{a})}$:
\begin{equation}
    \mathcal{L}_{\textnormal{self}}^{(a \leftrightarrow \hat{a})}=\cfrac{1}{N}\sum_{i=1}^N (l_{i}^{(a\rightarrow \hat{a})} + l_{i}^{(\hat{a}\rightarrow a)})
    \label{loss:self}
\end{equation}
where $l_{i}^{(a\rightarrow \hat{a})}$ and $l_{i}^{(\hat{a}\rightarrow a)}$ are defined in a similar way as in Eq.~\ref{loss:mini_con} and \ref{loss:mini_con2}. This loss function is useful to learn subtle differences over sound inputs as it needs to maximize the mutual information between two different views of the same inputs but to minimize the mutual information between two views of the different inputs. For example, as shown in Fig.~\ref{fig:new_figure}, an audio sample $a_i$ forms a negative pair with $\hat{a}_j$ for $i\neq j$, which induces a diffusive effect in the embedding space.

\myparagraph{Data Augmentation.}
We further apply the data augmentation strategy to improve the quality of representations and to overcome the lack of large-scale audio-text multimodal datasets. For audio inputs, we apply the SpecAugment~\cite{park19e_interspeech}, which visually augments Mel-spectrogram acoustic features by warping the features and masking blocks of frequency channels. For text inputs, we augment text data by (i) replacing words with synonyms, (ii) applying a random permutation of words, and (iii) inserting random words. Note that, for (i) we find synonyms of the given word from WordNet~\cite{fellbaum2010wordnet} and insert the synonym anywhere randomly in the given text input. For example, we augment original texts {\it ``rowboat, canoe, kayak rowing"} to produce new text {\it ``row canoe, kayak quarrel rowboat."}


\myparagraph{Loss Function.}
To summarize, we minimize the following loss function $\mathcal{L}_\text{total}$:
\begin{equation}
    \begin{aligned}
    \mathcal{L}_{\textnormal{total}} = \mathcal{L}_{\textnormal{nce}}^{( a \leftrightarrow v)} + \mathcal{L}_{\textnormal{nce}}^{( a \leftrightarrow t)} + \mathcal{L}_{\textnormal{self}}^{( a \leftrightarrow \hat{a})}
    \end{aligned}
    \label{loss:con}
\end{equation}

\subsection{Sound-guided Image Manipulation}
After learning the multi-modal joint embedding space by minimizing Eq.~\ref{loss:con}, we use a direct latent code optimization method to manipulate the given image similar to StyleCLIP~\cite{Patashnik_2021_ICCV}. As shown in Fig.~\ref{fig:contrastivelearning} (b), our model minimizes the distance between a given source latent code and an audio-driven latent code in the learned joint embedding space to produce sound-guided manipulated images. Moreover, we propose a {\it Adaptive Layer Masking} technique, which adaptively manipulates the latent code. 

 
 
\myparagraph{Direct Latent Code Optimization.} We employ the direct latent code optimization for sound-guided image manipulation by solving the following optimization problem:
\begin{equation}
    \begin{aligned}
    \mathcal{L}_{man} =\  & \underset{w_a \in \mathcal{W}+}\argmin\;{d_{\textnormal{cosine}}(G(w_a),a)} +
    \lambda_{\textnormal{ID}}{\mathcal{L}_{ID}}(w_a) \\
     & \hspace{3.57 cm} + \lambda_{sim}||{g} \cdot {(w_a - w_s)||_2}\\ 
    \end{aligned}
    \label{loss:man}
\end{equation}
where a given source latent code $w_s\in\mathcal{W}$~(the
intermediate latent space in StyleGAN), audio-driven latent code $w_a\in\mathcal{W}+$. $\lambda_{sim}$ and $\lambda_{ID}$ are hyperparameters. $g$ is a trainable vector to mask the specific style layer adaptively.
$\mathcal{L}_{\textnormal{ID}}$ and $G$ are the identity loss and StyleGAN-based generator, respectively. The source latent code $w_s$ means the randomly generated latent code from $G$ or the latent code obtained from the existing input image through GAN inversion~\cite{richardson2021encoding, 10.1145/3450626.3459838}. With such an optimization scheme, we minimize the cosine distance $d_{\textnormal{cosine}}(G(w_a),a)$ between the embedding vectors of the manipulated image $G(w_a)$ and the audio input $a$.


\myparagraph{Identity Loss.}
The similarity to the input image is also controlled by the identity loss function $\mathcal{L}_{\textnormal{ID}}$, which is defined:
\begin{equation}
    \mathcal{L}_{\text{ID}}(w_a) = 1 - \langle R(G(w_s), R(G(w_a))) \rangle
\end{equation}
where $R$ is the pre-trained ArcFace~\cite{deng2019arcface} model for face recognition, thus this loss function minimizes the cosine similarity $\langle R(G(w_s), R(G(w_a))) \rangle$ between its arguments in the latent space of the ArcFace network. This allows manipulating facial expressions without changing the personal identity. Note that we disable the identity loss by setting $\lambda_{\textnormal{ID}}=0$ for all other image manipulations.


\myparagraph{Adaptive Layer Masking.} 
We control style changes with adaptive layer masking.
$L_2$ regularization is effective in keeping the image generated from the moved latent code from being different from the original~\cite{Patashnik_2021_ICCV}. However, StyleGAN's latent code has different properties per each layer, so different weights should be applied to each layer if the user-provided attribute changes. We use layerwise masking to keep compact content information within style latent code. 
In StyleGAN2~\cite{karras2020analyzing}, the latent code represents as $ w \in \mathbb{R}^{L \times D}$, where $L$ is the number of the network layers, and $D$ is the latent code's dimension size. We declare a parameter vector $g$ in $L$ dimension.  In latent optimization step, $g$ and $w$ are multiplied per layer. $g$ is iteratively updated, which adaptively manipulates the latent code.

\myparagraph{Sound and Text Multi-modal Style Mixing.} Multi-modal manipulation of audio and text is based on style mixing of StyleGAN. Different layers of $w$ latent code in StyleGAN represent different properties. 
Because audio and text share the same new multi-modal embedding space, selecting a specific layer of each latent code guided by audio and text can manipulate the image using properties of audio and text. 


\section{Experiments}



\myparagraph{Implementation Details.}
Following CLIP~\cite{radford2learning}, we use the Vision Transformer (ViT)~\cite{dosovitskiy2021an} for our image encoder and the Transformer~\cite{radford2019language} for our text encoder. Note that we use a pre-trained model from \cite{radford2learning}. For our audio encoder, we use ResNet50~\cite{hershey2017cnn} by following~\cite{hershey2017cnn}, where we employ the same output dimension, 512, as the image and text encoder. First, we convert audio inputs to Mel-spectrogram acoustic features. Then, our audio encoder takes these features as an input to produce a 512-dimensional latent representation. The details about the train dataset are explained in supplemental material. For the manipulation step, we leverage StyleGAN2~\cite{karras2020analyzing}'s pre-trained generator. We set the size of latent code based on the resolution of the learned image. Here, we set $18 \times 512$ for images of size $1024 \times 1024$ and $14 \times 512$ for $256 \times 256$.

We train our model for 50 epochs using the Stochastic Gradient Descent (SGD) with the cosine cyclic learning rate scheduler~\cite{smith2017cyclical}. We set the learning rate to $10^{-3}$ with the momentum $0.9$ and weight decay $10^{-4}$. The batch size is set to 384. For audio augmentation, we use SpecAugment~\cite{park19e_interspeech} with the frequency mask ratio of $0.15$ and time masking ratio of $0.3$. For direct latent code optimization, $\lambda_{sim}$ and $\lambda_{ID}$ in in Eq. (\ref{loss:man}) are set to $0.008$ and $0.004$ for the FFHQ dataset; and $0.002$ and $0$ for the LSUN dataset.



\begin{figure}[h]
  \centering
  \includegraphics[width=\linewidth]{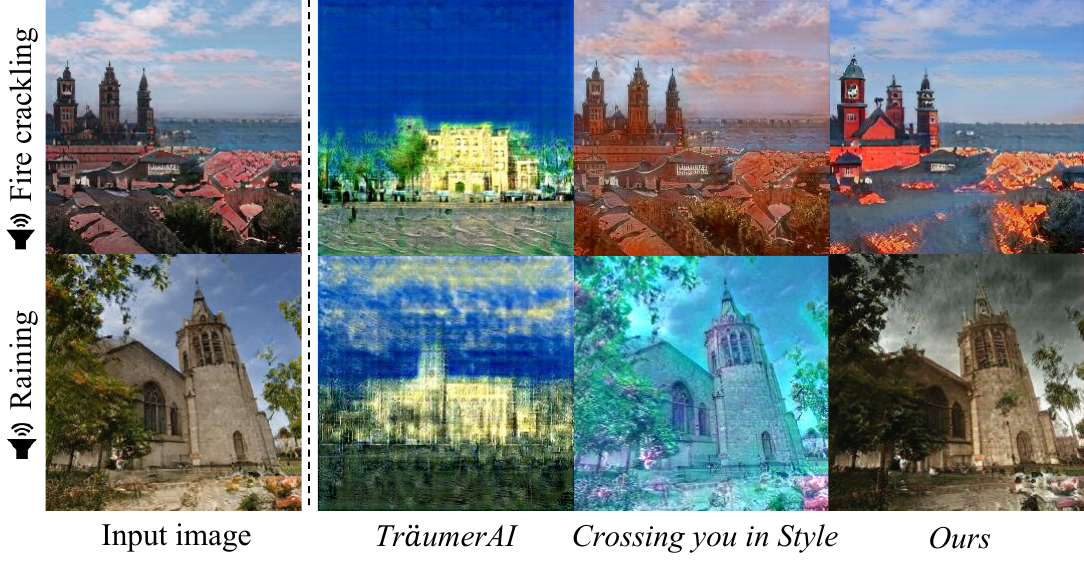}
  \vspace{-1.5em}
  \caption{Comparison of sound-guided manipulation results. Given fire crackling (top) and raining (bottom) audio inputs, we manipulate the input image with Tr\"{a}umerAI~\cite{jeong2021tr}, Crossing you in style~\cite{lee2020crossing}, and our method.}
  \vspace{-1.5em}
  \label{fig:cvprfig5}
\end{figure}

\subsection{Qualitative Analysis}


\myparagraph{Sound-guided Image Manipulation.} 
We first compare our sound-guided image manipulation model with existing sound-based style-transfer models including Tr\"{a}umerAI~\cite{jeong2021tr} and Crossing you in Style~\cite{lee2020crossing}. Fig.~\ref{fig:cvprfig5} showcases image manipulation results in response to given audio inputs, including fire crackling and raining. We observe that our model produces a better quality of manipulated images where existing models often fail to capture semantic information of the given audio input~(See 2\textsuperscript{nd} and 3\textsuperscript{rd} columns). 



\begin{figure}[t]
  \centering
  \includegraphics[width=\linewidth]{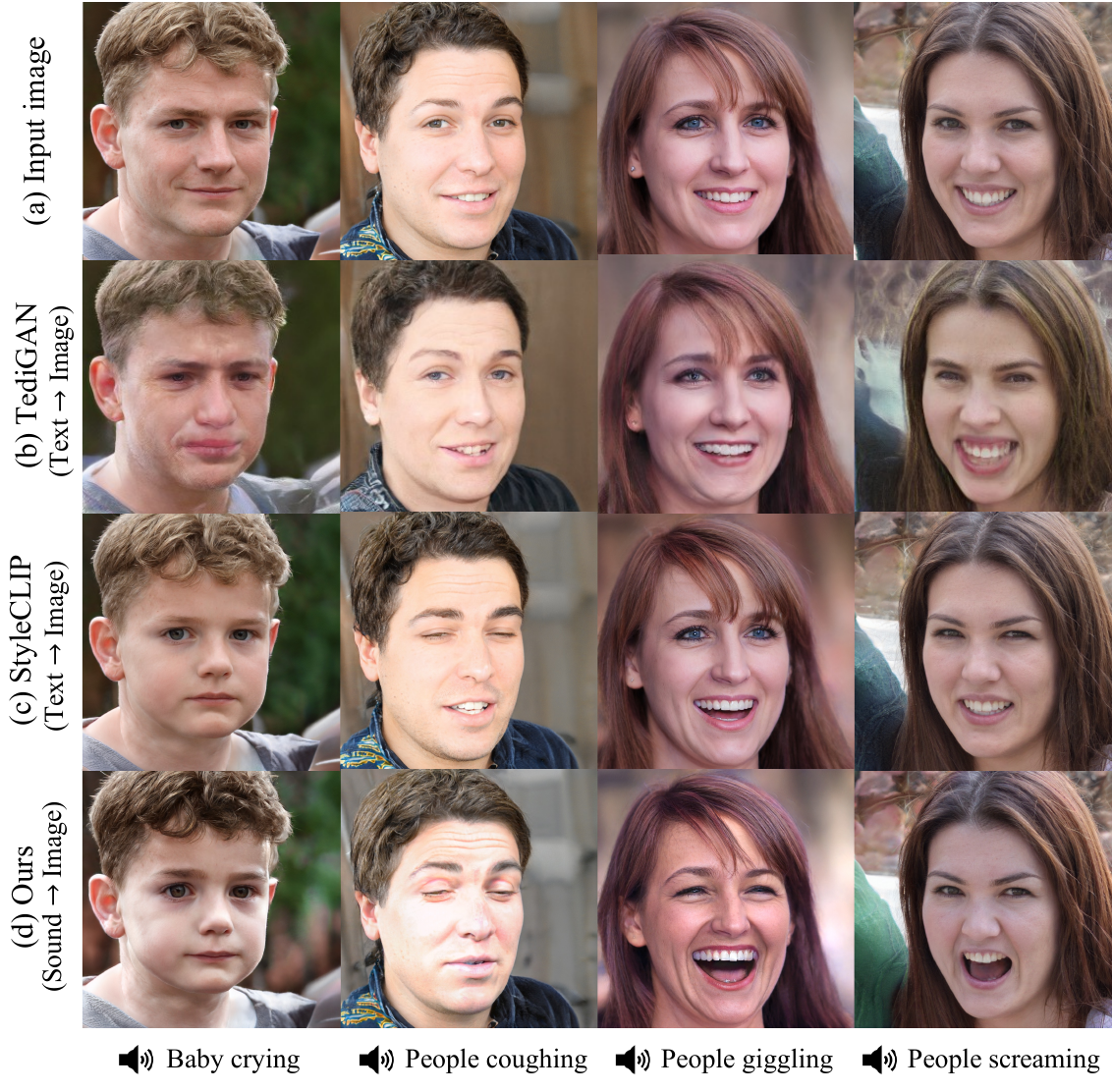}
  \vspace{-1.5em}
  \caption{Given the (a) input image, we compare the image manipulation results between (b-c) text-driven image manipulation approaches (i.e. TediGAN~\cite{xia2021tedigan} and StyleCLIP~\cite{Patashnik_2021_ICCV}) and (d) ours. Attributes for driving such manipulations include baby crying, people coughing, people giggling, and people screaming.}
  \label{fig:cvprfig2}
  \vspace{-1.5em}
\end{figure}

\myparagraph{Comparison of Text-guided Image Manipulation.}
We use the latest text-guided image manipulation models as a baseline, including TediGAN and the latent optimization technology of StyleCLIP. As shown in Fig.~\ref{fig:cvprfig2}, the proposed sound-guided image manipulation~(proposed method) shows more radical results than text-guided manipulation~(TediGAN~\cite{xia2021tedigan} and StyleCLIP~\cite{Patashnik_2021_ICCV}). Unlike text-guided methods, the audio-guided approach achieves natural image style transfer while capable of reflecting multiple labels.
 For example, TediGAN emphasizes crying, whereas StyleCLIP focuses on the baby when ``baby crying'' context is given. On the contrary, our proposed method is capable of handling ``baby'' and ``crying'' simultaneously. 

\begin{figure}[t!]
  \centering
  \includegraphics[width=\linewidth]{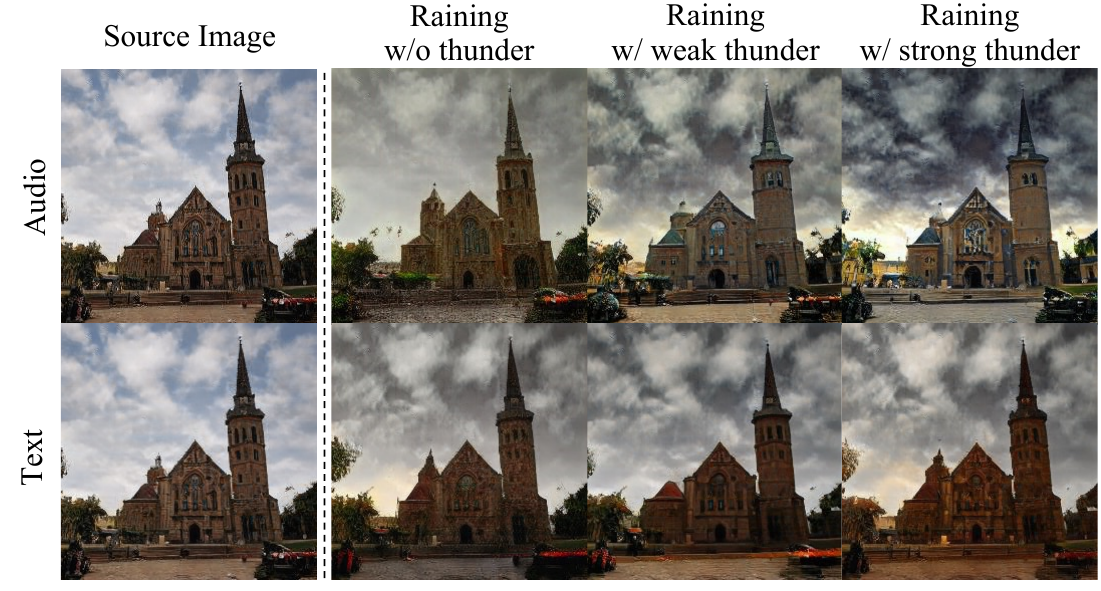}
  \vspace{-1.3em}
  \caption{Comparison of manipulation results between ours (top) and the existing text-driven manipulation approach, StyleCLIP~\cite{Patashnik_2021_ICCV} (bottom). Unlike the text-driven approach, ours can produce more diverse manipulation results in response to different intensities of raining, i.e. raining, raining with weak thunder, and raining with strong thunder.}
  \label{fig:difference}
  \vspace{-0.8em}
\end{figure}

We demonstrate that each audio sample has its own context, which makes the guidance richer than text~(Fig.~\ref{fig:difference}). If the magnitude of \textit{Thunder} is altered or a specific attribute like \textit{Rain} is added to the audio, the manipulation context becomes more diverse than text-guided image manipulation.

We visualize the direction vector with t-SNE~\cite{van2008visualizing} in a supplemental document. By subtracting the vectors of the latent code guided by each modality and the source latent code, we show the distribution of manipulating direction. 
We select the attributes in VGG-Sound~\cite{chen2020vggsound} and randomly manipulate the audio and text prompts.
Although we randomly sample the audio and text in the same labels, the sound-guided latent code shows a more significant transition than the text-guided latent code. We use various text synonyms for a fair comparison, but text-guided latent code seems less effective with changes. 

\begin{figure}[t!]
  \centering
  \includegraphics[width=\linewidth]{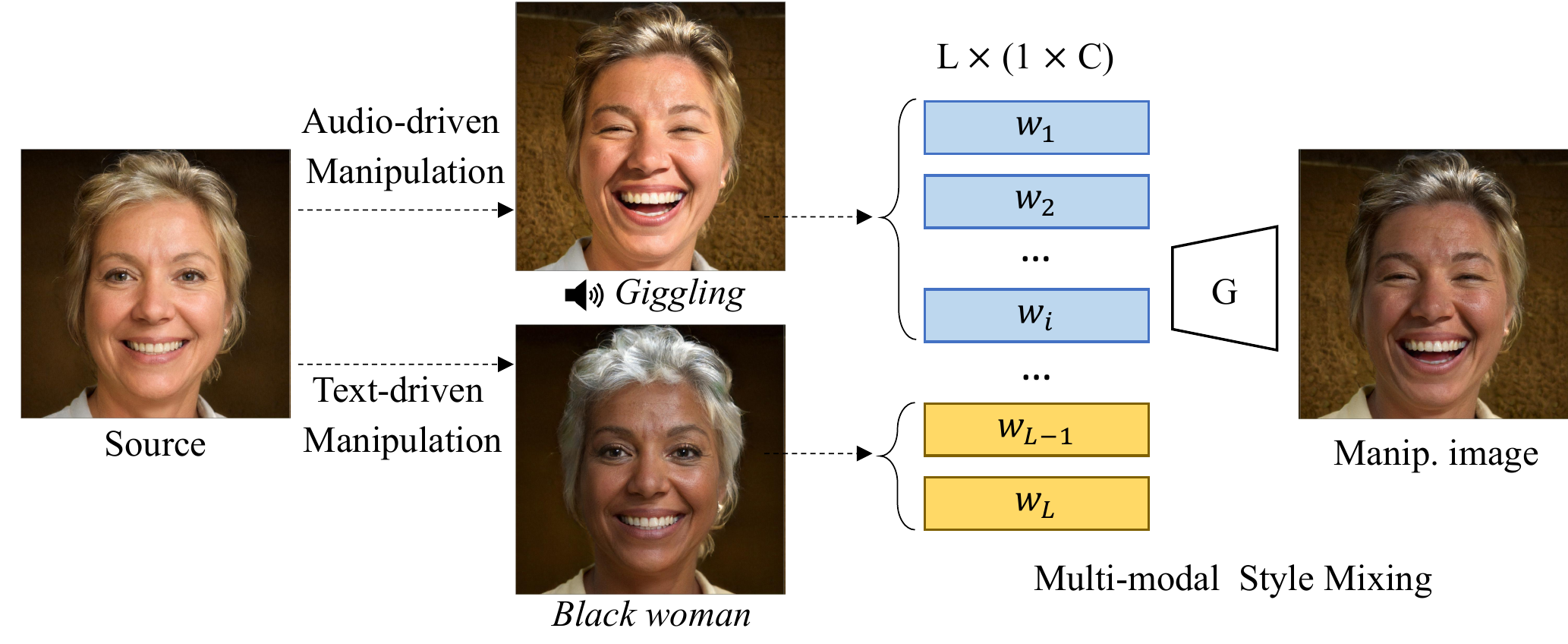}
  \caption{An example of  image style mixing jointly with the audio (\textit{people giggling}) and text input (\textit{black woman}).}
  \label{fig:stylemixing}
   \vspace{-2em}
\end{figure}

\myparagraph{Multi-modal Image Manipulation.} 
Our method ensures that audio, text, and image share the same embedding space. To demonstrate that multi-modal embedding lies in a same latent space, we interpolated text and sound-guided latent code~(see supplementary document). 
Constructing multi-modal shareable latent space enables joint modification of the target image
with user-provided text and audio inputs from the same embedding space.
We further perform multi-modal style mixing experiments by selecting a specific layer of latent code and mixing style with audio and text. 
We find that the sound source can effectively manipulate facial emotion aspects such as ``giggling" on the face and text information controls the background color of the target image (Fig.~\ref{fig:stylemixing}). 
For the style-mixing details, we follow TediGAN's StyleGAN layerwise analysis~\cite{xia2021tedigan}. In the 18 $\times$ 512 latent code, the style-mixing technique selects the 1st to 9\textsuperscript{th} layers of the sound-guided latent code and the 10\textsuperscript{th} to 18\textsuperscript{th} layers of the text-guided latent code to mix the dynamic characteristics of sound and human properties of text.

\begin{figure}[t]
  \centering
  \includegraphics[width=\linewidth]{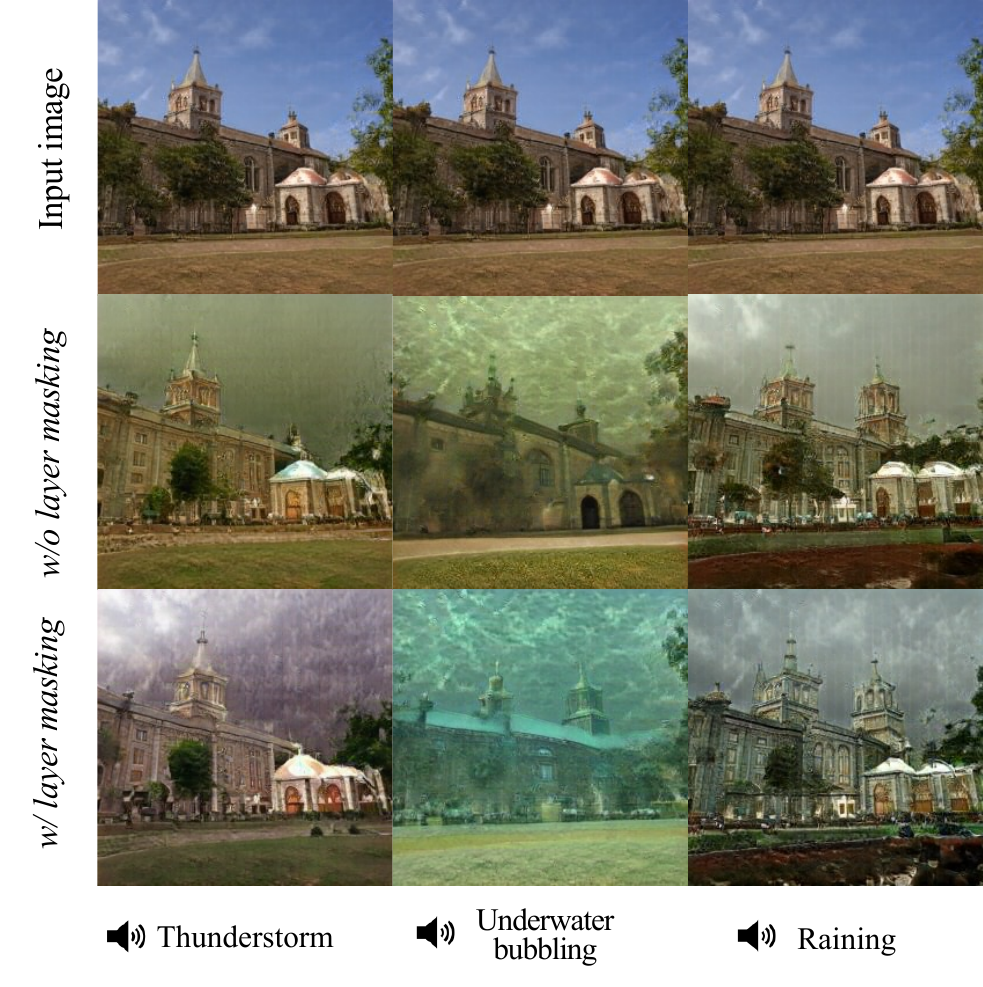}
  \vspace{-1.3em}
  \caption{Ablation study of adaptive layer masking. The first row is the input image, the second row is the manipulation result when the gate function is not applied, and the third row is the sound-guided image manipulation result after the gate function is applied. }
  \label{fig:cvprfig6}
  \vspace{-1.3em}
\end{figure}

\myparagraph{Effect of Adaptive Layer Masking.}
In StyleGAN~\cite{jeong2021tr}, it is necessary to adaptively regularize style layer since each layer of latent code has different style attributes. For each layer of latent code, it multiplies the trainable parameter that controls the diversity
during regularization. 
The ablation study shows a qualitative comparison of the mechanism for applying adaptive layer masking to the style layer, as illustrated in Fig.~\ref{fig:cvprfig6}. The adaptive masking rectifies the direction by changing the latent code based on the semantic cue. When applying the gate function, sound-guided image manipulation is semantically reasonable. For example, a thunderstorm is a blend of thunder and rain sound. Although thunder and lightning are not seen in the second row, lightning and rain appear in the last row. Manipulation results according to $\lambda_{sim}$ and $\lambda_{ID}$ hyperparameters are added to the supplemental material.

\subsection{Quantitative Evaluation}
\myparagraph{Zero-shot Transfer.} 
We compare our model to the supervised method and the existing zero-shot audio classification method. First, we compare audio embeddings trained by supervised methods such as logistic regression, ResNet50~\cite{hershey2017cnn} supervised by random initialization of weights as a baseline model, and AudioCLIP~\cite{guzhov2021audioclip}. We consider AudioCLIP as supervised learning method since it fine-tunes the evaluation dataset using the audio head in the paper. Even though logistic regression is used without additional fine-tuning on the ResNet50 backbone, it is comparable to AudioCLIP using ESResNeXt~\cite{guzhov2021esresne} as the backbone. Secondly, we compare the zero-shot audio classification accuracy with Wav2clip~\cite{wu2021wav2clip}. Table~\ref{zero} shows that our model outperforms previous studies in each task. Our proposed loss learns three modalities in the CLIP embedding space and obtains a more rich audio representation through the contrastive loss between audio samples whereas Wav2clip only learns the relationship between audio and visual.

\begin{figure}[t]
  \centering
  \includegraphics[width=\linewidth]{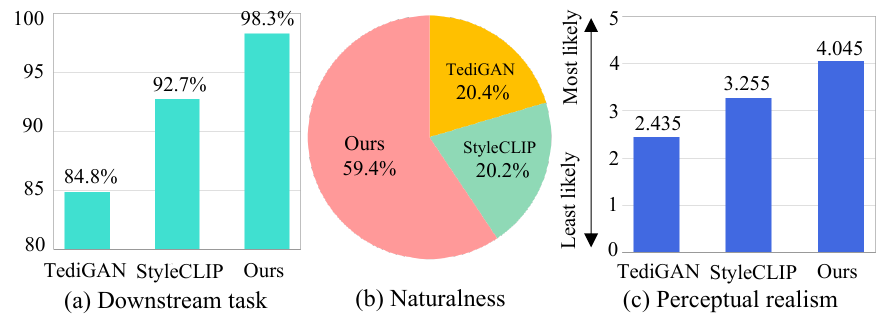}
  \caption{Quantitative evaluation and user study results. (a)~Downstream task evaluation to compare the quality of representations between ours and text-driven manipulation approaches on the FFHQ~\cite{karras2019style} dataset. A linear classifier is trained to predict 8 semantic labels, such as giggling, sobbing, etc. Participants answered a questionnaire including (b)~naturalness (``Which of the images is the best?'') and (c)~perceptual realism (``Do you think the provided image looks naturally manipulated?''). For perceptual realism, we use a 5-point Likert scale.}
  \label{fig:cvprfig13}
  \label{fig:userstudy}
  \vspace{-1.3em}
\end{figure}

\myparagraph{Semantic Accuracy of Manipulation.} We quantitatively analyze the effectiveness of our proposed audio-driven image manipulation approach. First, we measure performance on the semantic-level classification task. Given the audio embeddings from our pre-trained audio encoder, we train a linear classifier to recognize eight semantic labels including giggling, sobbing, nose-blowing, fire crackling, wind noise, underwater bubbling, explosion, and thunderstorm. We use StyleGAN2~\cite{karras2020analyzing} weights pre-trained from the FFHQ~\cite{karras2019style} dataset when guiding with giggling, sobbing, and nose blowing attributes to compare the semantic-level classification accuracy between text and audio. Also, when guiding with fire crackling, wind noise, underwater bubbling, explosion, and thunderstorm attributes, the weights of StyleGAN2 pre-trained with the LSUN (church)~\cite{yu2015lsun} dataset are used. As shown in Fig.~\ref{fig:userstudy}~(a), we generally outperform existing text-driven manipulation approach with better semantically-rich latent representation. 

\begin{table}[t!]
    \caption{Comparison of the quality of audio representations between ours and alternatives. We report classification accuracy (top-1 in \%) of a linear classifier on the ESC-50~\cite{piczak2015esc} and the Urban sound 8k~\cite{Salamon:UrbanSound:ACMMM:14} datasets as well as their zero-shot inference results. {\it{Abbr.}} $S$: supervised setting.}
    \label{zero}
    \centering
    \resizebox{\linewidth}{!}{
    \begin{tabular}{@{}lcccc@{}}
        \toprule
        \multirow{2}{*}{Model} & \multirow{2}{*}{$S$} & \multirow{2}{*}{Zero-shot} & \multicolumn{2}{c}{Dataset} \\ \cmidrule{4-5}
        & & & ESC-50 & Urban sound 8k \\ \midrule
        ResNet50~\cite{hershey2017cnn} & \checkmark & - & 66.8 \% & \textbf{71.3 \%} \\ 
        AudioCLIP~\cite{guzhov2021audioclip} & \checkmark & - & 69.4 \% & 68.8 \% \\ \midrule
        Ours w/o $\mathcal{L}_{nce}^{(a \leftrightarrow \hat{a})}$ & - & - & 58.7 \% & 63.3 \% \\
        Ours & - & - & \textbf{72.2 \%} & 66.8 \% \\\midrule
        Wav2clip~\cite{wu2021wav2clip} & - & \checkmark & 41.4 \% & 40.4 \% \\
        Ours w/o $\mathcal{L}_{nce}^{(a \leftrightarrow \hat{a})}$ & - & \checkmark & 49.4 \% & 45.6 \% \\
        Ours & - & \checkmark & \textbf{57.8 \%} & \textbf{45.7\%} \\
        \bottomrule
    \end{tabular}}
    \vspace{-0.5em}
\end{table}


\myparagraph{Distribution of Manipulation Direction.} We can see how much the latent code has changed by the cosine similarity between the source latent code and the manipulated latent code. We compare the cosine similarity between text-guided and sound-guided latent representations. We evaluate the mean and variance of the cosine similarity between $w_s$, a source latent code, $w_a$, an audio-driven latent code, and $w_t$, a text-driven latent code. The latent representations generally exhibit a high-level characteristic of the content~(see sup.). In the latent space of StyleGAN2, the sound-guided latent code moves more from the source latent code than the text-guided latent code, and the image generated from the sound-guided latent code is more diverse and dramatic than the text-guided method.

\subsection{User Study}
We recruit 100 participants from Amazon Mechanical Turk (AMT) for evaluating our proposed method. 
We show participants three manipulated images that are generated by TediGAN~\cite{xia2021tedigan}, StyleCLIP~\cite{Patashnik_2021_ICCV}, and our model. Participants answer the following questionnaire: (i) Perceptual Realism-~\textit{Which of the images is the best?} and (ii) Naturalness-~\textit{Do you think the provided image looks naturally manipulated?} For naturalness, we employ Likert scale ranging from 1~(low naturalness) to 5~(high naturalness). Fig.~\ref{fig:userstudy}~(b) and Fig.~\ref{fig:userstudy}~(c) show that our method significantly outperforms other state-of-the-art approaches~(TediGAN and StyleCLIP) in terms of \textit{Perceptual Realism} and \textit{Naturalness}. The large portion of participants~($59.4$\%) chose generated image by our model as the best. Moreover, the result also shows that our method generated more natural images than other text-driven manipulation approaches. Details are illustrated in supplementary document.

\section{Applications}
\begin{figure}[t]
  \centering
  \includegraphics[width=\linewidth]{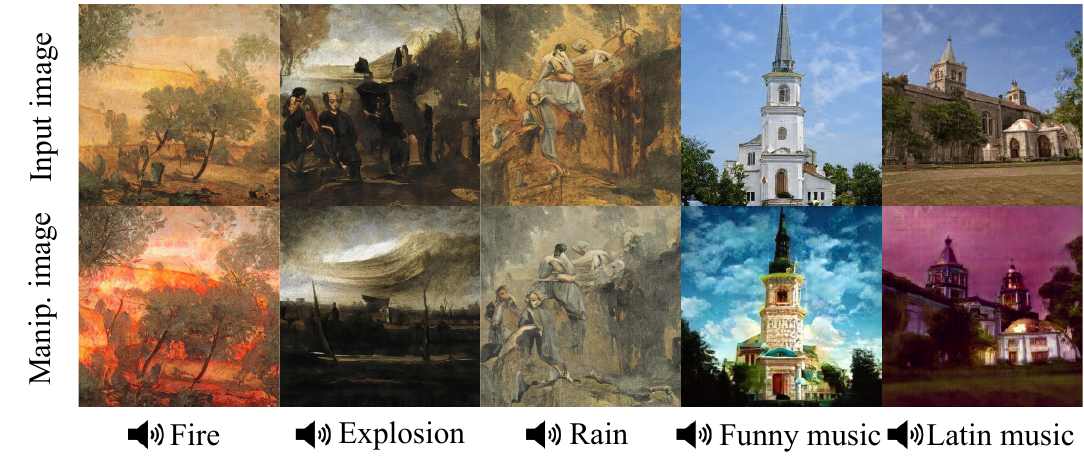}
  \caption{Examples of sound-guided artistic paintings manipulation and music style transfer using our method.}
  \label{fig:cvprfig8}
  \vspace{-1.3em}
\end{figure}

\myparagraph{Sound-Guided Artistic Paintings Manipulation.} 
We propose a novel sound-guided image manipulation approach for artistic paintings. We employ StyleGAN2~\cite{karras2020analyzing} generator which is pre-trained with the fine-art paintings dataset called WikiArt~\cite{saleh2016large}. As shown in Fig.~\ref{fig:cvprfig8}, our model could produce various manipulations for art paintings guided by given audio inputs. We observe that an audio input can successfully provide a semantic cue to manipulate artistic paintings. Given a fire crackling sound, a painting is manipulated with fire crackling. We also measured the manipulation quality for artistic painting using the Wikiart dataset using AMT. The responses showed that audio~(73.3\%) is better than text~(26.7\%) in terms of manipulation.

\myparagraph{Music Style Transfer.}
Our method has a potential to reflect the mood of the music into the image style. Fig.~\ref{fig:cvprfig8} illustrates the results of image style transfer with various music genres. The source latent code is close to the keywords of each music, so the mood of the music appears in the image. For instance, \textit{Funny} music manipulates the image with a fairy-tale style whereas \textit{Latin} music manipulates the image with red-color theme which reflects \textit{passion} characteristic. 

\begin{figure}[h]
  \centering
  \includegraphics[width=\linewidth]{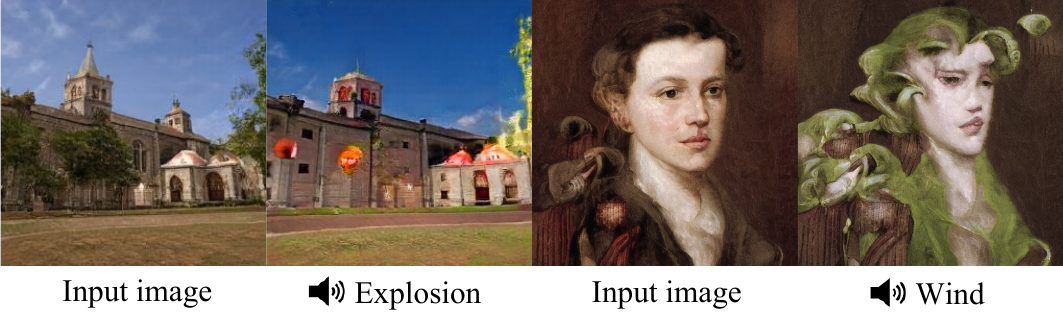}
  \vspace{-1.3em}
  \caption{Failure cases of manipulation with our method.}
  \label{fig:cvprfig7}
  \vspace{-1.3em}
\end{figure}

\section{Discussion and Conclusion}
We propose a method to manipulate images based on the semantic-level understanding from the given audio input. 
We take the user-provided audio input into the latent space of StyleGAN2~\cite{karras2020analyzing} and the CLIP~\cite{radford2learning} embedding space. Then, the latent code is aligned with the audio to enable meaningful image manipulation while reflecting the context from the audio. 
Our model produces responsive manipulations based on various audio inputs such as wind, fire, explosion, thunderstorm, rain, giggling, and nose blowing. We observe that an audio input can successfully provide a semantic cue to manipulate images accordingly. However, it would be challenging to preserve the identity for all cases due to the drastic change in image style~(see Fig.~\ref{fig:cvprfig7}).
Our method of traversing multi-modal embedding space can be used in many applications with multi-modal contexts.


\newpage
{
\small
\bibliographystyle{ieee}
\bibliography{egbib}
}

\end{document}